\documentclass[12pt,a4paper]{article}

\usepackage{array}
\usepackage{graphicx}
\usepackage{amsfonts}
\usepackage{amssymb,amsmath,amsthm}

\newcommand{\be}{\begin{eqnarray}}
\newcommand{\ee}{\end{eqnarray}}
\newcommand{\mycaption}[1]{\caption{{\em #1}}}

\renewcommand{\d}{\mathrm{d}}   

\renewcommand{\Re}{\mathfrak{Re}}

\renewcommand{\equiv}{{\; := \;}}

\newcommand{\Aut}{\mathrm{Aut}}
\newcommand{\cc}{\mathrm{c}}
\newcommand{\cf}{{\em cf.}}
\newcommand{\Cset}{\mathbb{C}}
\newcommand{\Der}{\mathrm{Der}}

\newcommand{\Det}{\mathrm{Det}}
\newcommand{\Dset}{\mathbb{D}}

\newcommand{\e}{{\mathrm e}}
\newcommand{\eg}{{\em e.g.}}
\newcommand{\Eset}{\mathbb{E}}

\newcommand{\Hset}{\mathbb{H}}
\newcommand{\ie}{{\em i.e.}}

\newcommand{\Lie}{\mathrm{Lie}}

\newcommand{\Obdle}{\mathfrak{L}} 
 
\newcommand{\Prob}{\mathfrak{P}}

\newcommand{\Pexp}{{{\cal P}\mathrm{exp}}}

\newcommand{\Rset}{\mathbb{R}}

\newcommand{\Tset}{\mathbb{T}}

\newcommand{\unit}{\mathbf{1}}
\newcommand{\Vir}{\mathrm{Vir}}

\newcommand{\Wt}{{W_t}}
\newcommand{\Xdet}{{|{\det}_X|}}
\newcommand{\Xpdet}{{|{\det}_{X,p}|}}
\newcommand{\Xope}{\mathfrak{O}}
\newcommand{\Zset}{\mathbb{Z}}

\begin{document}
\begin{titlepage}
\begin{flushleft}
\hfill IHES/P/03/28 \\
\hfill Imperial/TP/2-03/30 \\
\hfill {\tt hep-th/0308020}
\end{flushleft}

\vspace*{7mm}

\begin{center}
{\bf \Large  On Conformal Field Theory and \\
Stochastic Loewner Evolution} \\
\vspace*{8mm}

{      R.~Friedrich$^\mathrm{ab}$ and
       J.~Kalkkinen$^\mathrm{c}$
} \\

\vspace*{3mm}

${}^\mathrm{a}$
{\em Institut des Hautes \'Etudes Scientifiques, Le Bois-Marie} \\
{\em 35, Route de Chartres, Bures-sur-Yvette F--91440, France} \\
\vspace{2mm}
${}^\mathrm{b}$
{\em Universit\'e Paris-Sud, Laboratoire de Math\'ematiques} \\
{\em Universit\'e Paris XI, F--91504 Orsay, France} \\
\vspace{2mm}
${}^\mathrm{c}$ {\em The Blackett Laboratory, Imperial College} \\
{\em Prince Consort Road, London SW7 2BZ, U.K.} \\

\vspace*{6mm}

\end{center}

\begin{abstract}
We describe Stochastic Loewner Evolution on arbitrary Riemann
surfaces with boundary using Conformal Field Theory methods. We
propose in particular a CFT construction for a probability measure
on (clouded) paths, and check it against known restriction
properties. The probability measure can be thought of as a section
of the determinant bundle over moduli spaces of Riemann surfaces.
Loewner evolutions have a natural description in terms of random walk
in the moduli space, and the stochastic diffusion equation
translates to the Virasoro action of a certain weight-two operator
on a uniformised version of the determinant bundle.

\vfill

\begin{tabular}{ll}
{\em PACS 2003:} &
02.50.Ey, 
05.50.+q, 
11.25.Hf  
\\
{\em MSC 2000:}   &
60D05, 
58J52, 
58J65, 
81T40 
\\
{\em Keywords:}  & Probability Theory; Conformal Field Theory \\
 & \\
{\em Email:}    &  {\tt rolandf@ihes.fr, jek@imperial.ac.uk}
\end{tabular}
\end{abstract}
\end{titlepage}

\section{Introduction}
\label{intro}

The motivation for this article stems from the work of Lawler,
Schramm and Werner \cite{LSW} in which they investigate on a
purely probabilistic basis the ``restriction property'' of certain
probability amplitudes. This property can be phrased in terms of a
class of stochastic processes defined on proper simply connected
domains of the complex plane, originally introduced in
Ref.~\cite{S}. These processes -- termed Stochastic (or
Schramm--)Loewner Evolutions (SLE) -- proved to be of great value
in providing a rigorous basis for certain results in Conformal
Field Theory (CFT) \cite{LSW1, LSW2, LSW3, SS2}.

The precise relation of SLEs to CFT has remained, however,  rather 
speculative. Some aspects of it have been clarified building on the restriction property in Refs.~\cite{FW1, FW2}, where an
interpretation of the parameters involved in the ``Brownian bubble 
process" as the central charge and the conformal weight of a highest weight representation of an underlying CFT was obtained. This led further to an explicit calculation of the correlation functions of the stress--energy tensor inserted at the boundary for trivial central charge $c=0$. In this paper we continue to investigate the properties of the objects found in \cite{LSW} and propose a precise CFT model for 
chordal SLE that to some extent generalises it. In a forthcoming paper \cite{FK:V} we will expand on the mathematical side of this
article. An introduction to the present literature and to
remaining challenges in particular in the case of percolation can
be found in \cite{Cardy:2001vr,WW}.

Recall that a class of statistical mechanics models at the critical temperature can be described in terms of CFT. 
We will investigate models defined on domains with boundaries, or more generally on surfaces with boundaries (open string world-sheets). The choice of appropriate boundary conditions will give rise to a long chordal domain wall, connecting two points of the (same) boundary component. Since these fluctuating curves exist on all lengths scales it is  meaningful to ask for the  limiting object: In the case of  simply connected planar domains it has been shown \cite{S} that the scaling limit corresponds to  a class of  conformally invariant probability measures, supported by these random cruves. We will show here how to relate, in the general case, such random curves to the vacuum expectation values of certain nonlocal operators inserted on the (boundaries of) the manifold. The proper treatment of random curves starting from a point in the bulk and connecting another point on the boundary is related to order disorder lines which we will not treat in the present article.

In the general situation we will show how specific densities associated to  chordal  curves have the same covariance  properties under conditioning  as are known in the case
of chordal SLE on the upper half-plane.


The diffusion equation, which describes the stochastic process,
can be thought of as a real Schr\"odinger equation of a quantum
mechanical particle on the boundary. The Hamiltonian, which is a
second-order hypo-elliptic differential operator, is the generator
of the stochastic process. At zero energy, that is to say when the
Hamiltonian annihilates the wave function, the probability density
is a martingale.

The probability densities of SLE are conformal forms on the moduli
space of Riemann surfaces with boundary components and one marked
point ${\cal M}_{g,b,1}$. They can be assumed to belong to highest weight representations of the Virasoro algebra of the CFT, the weight itself being determined by the variance of the driving
Brownian motion \ie~the diffusion coefficient. From this point of
view the martingale property corresponds to requiring that the
Hamiltonian shifts the conformal weight of the wave function by
two, and that the resulting representation is still of highest
weight, and singular, so that it is equivalent to zero in the
physical Hilbert space. The fact that representations of the
Virasoro algebra were involved in the description of SLE curves
was noticed already in \cite{FW1,Bauer:2003di,FW2}.

As the Loewner process changes the conformal class of the Riemann
surface, it also generates a random walk in the moduli space  ${\cal
M}_{g,b,1}$. We describe this in detail, and write down the
probability densities as sections of certain line
bundles on this space. We show in particular that the induced motion 
on the moduli space has a tangent vector field that lies entirely
in a rank-two vector bundle determined by the second-order
differential operator.

Suppose we are given a differentiable path $\gamma_t$ in the upper
half-plane $\Hset$. 
Then we can make a polygonal approximation of  it with (infinitesimal) straight lines. The Riemann mapping theorem guarantees the existence of a conformal transformation that maps the upper half-plane minus a first part of the Jordan curve onto the upper half-plane. This map is, after a suitable normalisation, unique. A particularly important question is its boundary behaviour. In our situation Carath\'eodory's extension theorem tells us that it extends continuously to the two-sided slit.


To illustrate the situation we give the inverse mapping for a slit 
$\ell$ of length $t\alpha^{\alpha}/(1-\alpha)^{\alpha}$, extending 
in from 0 at an angle $\pi \alpha
\in ]0,\pi[$
\be
g_t^{-1} : \Hset \longrightarrow \Hset \setminus\ell ~;~ z \mapsto
\left(z + t \right)^{1-\alpha} \cdot 
\left(z - \frac{\alpha}{1-\alpha} t \right)^{\alpha} ~. \label{HL}
\ee
The points $-t$ and $+\alpha t/(1-\alpha)$ are mapped to $0$, and $0$ 
itself to the tip at $t\alpha^{\alpha}/(1-\alpha)^{\alpha} ~ 
\e^{i\pi\alpha}$; this function is now normalised
so that $g_t(z) \longrightarrow z$ when $z$ tends to infinity. 
The exact location of the two zeroes does not affect the present
discussion.

Suppose we are given locally on the upper half-plane 
a conformally flat metric, say $\e^{\sigma}|\d z|^{2}$. Under the above coordinate transformation $z \mapsto f(z,\bar{z})$ the conformal class -- and hence the complex structure -- of a locally conformally flat metric changes according to $|\d z|^2 \longrightarrow |\d z + \mu \d \bar{z}|^2$. (We omit here the arising Weyl factors.) The thereto associated {\em Beltrami
differential} $\mu(z,\bar{z})$ can be used to detect where the
transformation ceases to be conformal, and by what amount. Given
the transformation $f=g_t^{-1}$ on $X$ it can be readily
calculated from the defining equation $\partial f = \mu
\bar{\partial}f$. The Beltrami differential associated to $g_t$
given in Eq.~(\ref{HL}) can be argued to be a 
distribution of the form
\be
\mu_t(z,\bar{z}) &=& -2\pi i ~ \frac{\alpha}{1-\alpha} (z-t)^2
\delta^{(2)}(z-t) ~, \label{lemBeltrami}
\ee
where the parameter $t$ is related to the length of the slit. This
distribution is, of course, trivial when integrated with regular
test functions. In our case it will appear with functions with
second-order poles precisely at $z=t$, so that it yields a finite
result as discussed in detail in Sec.~\ref{newSec}.

\vspace{5mm}

The plan of the paper is as follows: In Sections \ref{planelowner} 
and \ref{partition} we recall some facts about the probabilistic 
treatment of the chordal Loewner process on the upper half-plane, and 
partition functions of statistical mechanics and Conformal Field 
Theory, respectively. If not otherwise stated, we are always referring to the cordal case. In Section \ref{moduli} we describe the behaviour of CFT observables under the Loewner process, and 
propose a probability density associated to a specific path on 
the Riemann surface. In Section \ref{det} we rephrase this in 
terms of determinant bundles over moduli spaces of Riemann 
surfaces in order to give such a geometric description to 
the Loewner process in terms of these moduli spaces that the 
Virasoro action becomes apparent. In the discussion section 
\ref{disc} we draw the probabilistic and the CFT discussions together.

\section{Loewner process and restriction properties}
\label{planelowner}

Let us recall the definition of a chordal SLE${}_{\kappa}$  in the
upper half-plane $\Hset$ that starts from $0$ and continues to
infinity \cite{WW,L}:

Let $(\Omega,{\cal F}, ({\cal F}_t),P)$ be a standard filtered probability
space that is complete and continuous from the right. Given
$z\in\Hset, t\geq 0$, define $g_t(z)$ by $g_0(z)=z$ and
\be
\frac{\partial g_t(z)}{\partial t}  &=& \frac{2}{ g_t(z) - \Wt}
\label{lowner} ~.
\ee
Here $\Wt   \equiv \sqrt{\kappa}\, B_t$ is the standard
one-dimensional Brownian motion $B_t$, defined on
$\Rset_+\times\Omega$ with initial point 0 and with variance
$\kappa>0$. This means in particular that its
It\^{o}-differentials satisfy $(\d \Wt)^2 = {\kappa}~ \d t$.

Given the initial point $g_0(z)=z$, the ordinary differential
equation (\ref{lowner}) is well defined until a random time
$\tau_z$ when the right-hand side in (\ref{lowner}) has a pole.
Define the set $K_t$ in the closure of the upper half-plane as
$K_t \equiv \overline{\{z\in\Hset : \tau_z < t\}}$. 

The family $(K_t)_{t\geq0}$ is an increasing family of compact
sets, also called hulls, in the closed upper half-plane $\overline{\Hset}$ and $g_t$
is the uniformising map from $\Hset\setminus K_t$ onto $\Hset$. It
has been shown in \cite{RS, LSW2} that there exists a continuous process
$(\gamma_t)_{t\geq0}$ with values in $\overline{\Hset}$ such that
$\Hset\setminus K_t$ is the unbounded connected component of
$\Hset\setminus\gamma[0,t]$ with probability one. This process is
the trace of the SLE${}_{\kappa}$ and it can be recovered from
$g_t$, and therefore from $W_t$, by
\be
\gamma_t  &= & \lim_{z\rightarrow W_t, z\in\Hset} g_t^{-1}(z)~.
\ee
The constant $\kappa$ characterises the nature of the resulting
curves as classified in \cite{RS}. For $0 < \kappa \leq 4 $
SLE${}_{\kappa}$ traces over simple curves, for $4 < \kappa < 8 $
self-touching curves (curves with double points, but without
crossing its past) and, finally, if $8\leq \kappa $ the trace
becomes space filling.

Let  us consider the function $f_{t}(z) \equiv   g_t(z) - \Wt$
that satisfies by virtue of Eq.~(\ref{lowner}) the stochastic
differential equation
\be
\d f_{t}(z) &=&  \frac{2}{f_t(z)}\d t - \d \Wt \label{stdiff} ~.
\ee
Consider $k$ points $x_i$, $i=1,\ldots,k$  on the real axis $\Rset$ 
which we view as the boundary of the upper half plane $\partial \overline{\Hset}$, their images $y_i \equiv f_t(x_i)$  under the Loewner mapping, and a smooth function $F : {\Rset}^k \longrightarrow \Rset$ of the coordinates $y_i$, $i=1, \ldots, k$. Then $F_t \equiv F(f_t(x_1),...,f_t(x_k))$ is a new stochastic process, and It\^{o}'s formula and the ordinary differential equation (\ref{lowner}) yield, as in \cite{FW1},
\be
\d F_{t} &=& \d \Wt ~ {\cal L}_{-1} ~ F_{t} + \d t ~ \Big(
\frac{\kappa}{2} {\cal L}_{-1}^2 - 2  {\cal L}_{-2}  \Big) F_{t}
~, \label{dF}
\ee
where the differential operators ${\cal L}_n$ are given by
\be
{\cal L}_n &\equiv&  \sum_{j=1}^k  - y_j^{n+1}
\frac{\partial}{\partial y_j} ~.
\ee

Similarly one may construct slightly more general stochastic processes by pulling back holomorphic differential forms  on the upper half-plane $\Hset$ by the Loewner mapping $f_{t}$; the same algebra as above goes indeed through provided we restrict to holomorphic functions or, more generally, differential forms. The additional piece of information one requires is the fact that the standard de Rham differential $\d_{dR}$ acts in this situation as usual through 
\be
\d_{dR}\,  f_{t}(x_i) &=& \Big( f_t^* \d_{dR}\, y_i\Big)_{x_i} ~, 
\ee 
and that the It{\^o} differential $\d$ acts independently 
\be
\d \Big( \d_{dR}\,  f_t(x_i) \Big) &=&   \frac{ -2}{f_t(x_i)^2}\d
t \otimes \d_{dR}\,  f_t(x_i)~. 
\ee  
Consider in particular a product of holomorphic differential forms with conformal weights $h_{i}$, namely $\omega \in \bigotimes_i {\Omega^{(1,0)}(\Hset)}^{\otimes h_i}$. These objects are conformal forms evaluated at $k$ distinct points on $y_{i} \in \Hset$. 
Pulling back these forms with the Loewner mapping\footnote{Note that the operators ${\cal L}_n$ are expressed in terms of the coordinates $y_{i}$ and not of $x_{i}$.} $f_{t} : x_{i} \mapsto y_{i}$, we can define the stochastic process $\omega_t \equiv f_t^*\omega$. The components of the differential form $\omega_t$ obey still the equation (\ref{dF}), but the operators ${\cal L}_n$ are now defined as 
\be
{\cal L}_n &\equiv&  \sum_{j=1}^k  - y_j^{n+1}
\frac{\partial}{\partial y_j} + h_j(n+1) y_j^n \label{curlyL}
\ee
in terms of the local coordinates $y_{j}$ of the $j$th factor.  This is precisely the action of a generator of the Witt algebra, $\Der ~{\cal K}$, 
\be
[{\cal L}_n,{\cal L}_m] &=& (n-m) {\cal L}_{n+m} ~,
\ee
on a highest weight state, with conformal weights $h_{j}$. 
The Witt  algebra has a universal one-dimensional central extension 
\be
0 \longrightarrow \Cset \longrightarrow \Vir \longrightarrow  
\Der ~{\cal K} \longrightarrow 0~,
\label{cextension}
\ee
known as the Virasoro algebra $\Vir$. For a specific central
element $c \unit$, its generators $L_n$, $n \in \Zset$ obey the
following commutation relations
\be
[L_n,L_m] &=& (n-m) L_{n+m} + \frac{c}{12} n(n^2-1) \delta_{n+m}
{\bf 1} ~.
\ee
Later in the Paper, in particular in Sec.~\ref{det-loew},  we shall 
see the Witt algebra acting on certain objects and how this action can be extended to that of the Virasoro algebra. This will amount to replacing ${\cal L}_n$ by $L_n$. 

If we specialise now the above discussion to the case $k=1$ where 
$\omega\in\Omega^{(1,0)}(X)^{\otimes h}$ is a single holomorphic
conformal form on the Riemann surface $X = \Hset$ with conformal weight 
$h$ and $x \in X$ some choice of local coordinates on $X$, then  the 
stochastic process $\omega_t \equiv f_t^*\omega$ obeys the stochastic 
differential equation 
\be 
\d \omega_t &=& \d \Wt ~ L_{-1} ~ \omega_{t} + \d t ~ \Big(
\frac{\kappa}{2} {L}_{-1}^2 - 2  {L}_{-2}  \Big) \omega_{t}
\label{domega} ~.
\ee
This result implies in particular that the process
$\omega_t$ happens to be a martingale  for  SLE${}_{\kappa}$, \ie
\be
 \Big( \frac{\kappa}{2} {L}_{-1}^2 - 2  {L}_{-2}  \Big) \omega_{t}(x_i) =
 0 \label{singular}
\ee
precisely when 
\be
c &=& -\frac{(3\kappa-8)(\kappa-6)}{2\kappa} \label{ck} \\
h &=& \frac{6-\kappa}{2\kappa} \label{hk} ~. 
\ee
Then the descendants of $\omega$ do indeed contain the required 
singular (null) vector in the Verma module 
$V_{2,1}$.\footnote{We consider here the highest weight vector
submodule associated to this singular vector rather than the quotient of 
the full Verma module $V_{2,1}$ with it; The quotient construction 
with the values of $c$ and $h$ as in (\ref{ck}) and (\ref{hk}) has been 
considered independently from a different point of view in 
Refs.~\cite{Bauer:2002qn,Bauer:2002tf} for the upper half-plane.} 
These are also the only values of these parameters for which the 
drift term is a highest weight representation of the Virasoro algebra.

In the rest of the paper we shall be mostly interested in the behaviour of these differential form-valued stochastic processes on the boundary $\partial X$ of the Riemann surface $X$, which in this section is simply the upper half-plane. In this, as well as the more general case, the local coordinates $x \in \partial X$ on the boundary components are real. It is important, nevertheless, that the holomorphic analysis above goes through because this guarantees that we can express these boundary objects in terms of the right bulk objects as required in the Schottky double treatment of Boundary Conformal Field theory (BCFT). 
   
We make now use of the fundamental fact that one can associate a
second-order partial differential operator to It\^o diffusions,
namely its generator. For the stochastic differential equation
(\ref{stdiff}) we obtain
\be
\bar{{\cal H}} &=&  \frac{\kappa}{2} {\cal L}_{-1}^2 - 2  {\cal
L}_{-2} ~, \label{stochproc}
\ee
in the notation of definition (\ref{curlyL}) with $h=0$. 
The Feynman--Kac formula states that
if we define for $\varphi\in C_0^2(\Rset^n)$ and $V\in C^0(\Rset^n)$
bounded from above 
\be
\psi(x,t) & \equiv & \Eset^x \Big( \varphi(W_{t})\cdot \exp
\int^t_0 V(W_s) \d s \Big), \label{diffdistr}
\ee
then we have
\be
\frac{\partial\psi(x,t)}{\partial t} &=& \Big( \bar{{\cal H}}(x) +
V(x) \Big) \psi(x,t) \label{genfini} \\ \psi(x,0) &=& \varphi(x)
~,
\ee
where $\bar{\cal H}$ is the generator of the process $\omega_t$.\footnote{At this point we need to restrict to boundary points $x \in \partial X$, which are real.}
By choosing formally the potential function as
\be
 V(x) &  \equiv& \frac{2h}{x^2}~
\ee
we get the full Hamiltonian for arbitrary positive $h$
\be
\hat{\cal H} &\equiv&  \bar{{\cal H}}(x) + V(x) \\
 &=& \frac{\kappa}{2} {\cal L}_{-1}^2 - 2  {\cal L}_{-2}
\ee
as it appeared in Eq.~(\ref{stochproc}). This potential is of
course not bounded at the source $x=0$. Nevertheless, the
Fokker--Planck equation (\ref{genfini}) gives the time evolution
of the probability density $\psi(x,t)$, and can be interpreted as
a real Schr\"odinger equation. In that context $\psi$ is the wave
function of a quantum mechanical particle evaluated at time $t$
and at the initial position $x$.

\subsection{Restriction property}
\label{restriction}

Consider chordal SLE$_{\kappa}$ for $\kappa<4$ which produces a
simple curve  $\gamma : [0,\infty)\rightarrow\overline{\Hset}$
with $\gamma(0)=0$, $\gamma(0,\infty)\subset\Hset$, and
$\gamma(t)\rightarrow\infty$ as $t\rightarrow\infty$. Let the hull
$A$ be a compact set $A\subset\overline{\Hset}$ such  that
$A\cap\Rset\subset\Rset^{\ast}_+$, and $\Hset\setminus A$ is
simply connected. For such a hull  $A$,  let $\phi_A:
\Hset\setminus A\rightarrow\Hset$ be the conformal map that
preserves 0 and infinity such that $\phi'_A(\infty)=1$, \ie~that
it is hydrodynamically normalised. Under this normalisation
composition of uniformising maps defines a pseudo semi-group on
hulls: 
Suppose that $A$ and $A'$ are hulls 
and let $A_1 \equiv \Phi^{-1}_A(A')$. Then 
\be  
\phi_{A_1} &=& \phi_{A'}\circ\phi_A ~. 
\ee 

\begin{figure}[ht]
\begin{center}
\includegraphics[scale=0.6]{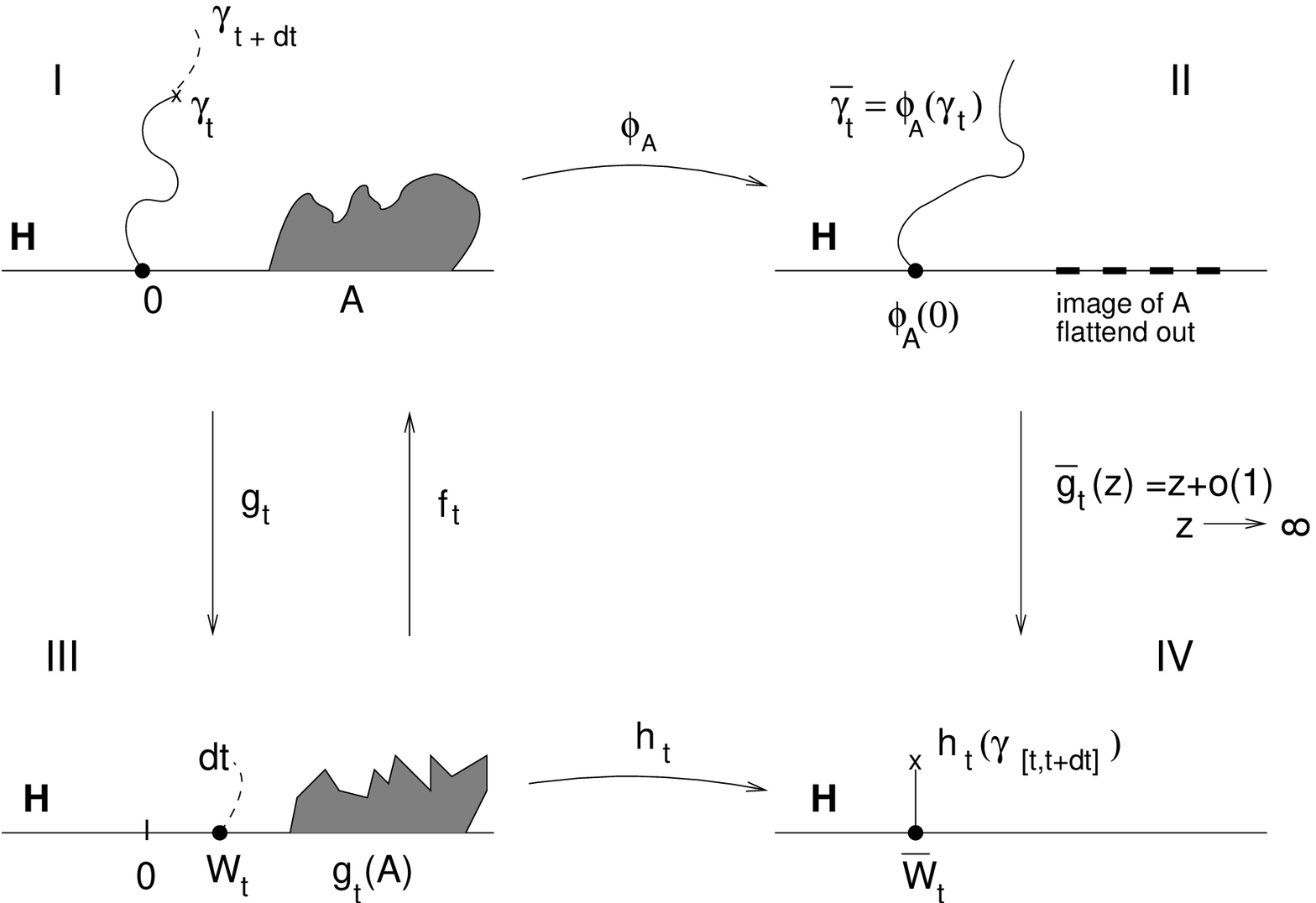}
\mycaption{The commutative diagram.} \label{cd}
\end{center}
\end{figure}

Let us further define the set $V_{\infty} \equiv\{\omega :
\gamma[0,\infty)\cap A=\emptyset\}$ which is measurable and has a
positive probability. Given a specific outcome $\gamma \in
V_{\infty}$ in this set, we may consider the path $\bar{\gamma}
\equiv\phi_A \circ \gamma(t)$. For all $A$ as above and assuming
$\gamma[0,\infty)\cap A=\emptyset$, we say that SLE$_{\kappa}$
satisfies the {\em chordal restriction property} if the
conditional distribution of $\bar{\gamma}$ is the same as a time
change of SLE$_{\kappa}$. This can be summarised in the
commutative diagram in Fig.~\ref{cd}. SLE$_{\kappa}$ has indeed
been shown to have this property for $\kappa = 8/3$ in
Ref.~\cite{LSW}.

On the other hand, restricting is the same as introducing a new
probability measure $Q_A$ under which all paths that meet the hull
$A$ in a finite time form  a set of measure zero. Using the
original probability measure $P$, the new measure $Q_A$ can be
simply defined as \be Q_A(G)&  \equiv&\frac{P(G\cap
V_{\infty})}{P(V_{\infty})}\quad\quad\mbox{where}\, G\in{\cal F}
\label{probcond} ~. \ee It is therefore a conditional probability.
Furthermore, it is absolutely continuous with respect to the
original measure $P$, \ie~$Q_A\ll P$ on ${\cal F}_{\infty}$. We
can therefore calculate the Radon--Nikod\'ym derivative of $Q_A$
with respect to $P$ on the sub-$\sigma$-algebra ${\cal F}_t$ and
express it in the form
\be
\label{girsanov} \d Q_A(\omega)|_{{\cal F}_t} &=& M_t(\omega) \d
P(\omega)|_{{\cal F}_t} ~.
\ee

This can actually also be extended to the range $\kappa \in
(0,8/3)$. This requires, however, surrounding the simple path
$\gamma$ with a cloud of Brownian bubbles $\Xi$ with a certain
intensity $\lambda$ as in Ref.~\cite{LSW}, and posing then the following 
question: What is the probability that the thus created hull around
the path $\gamma$ does not intersect the set $A$? The difference
between the original measure $P$ on $\gamma$ and the conditioned measure
on $\gamma$ can be expressed again through the Radon--Nikod\'ym derivative
\be
Y_t &\equiv& h_t'(\Wt)^\alpha ~ \exp \frac{\lambda}{6} \int_0^t
\Big\{h_s(W_s), W_s \Big\} ~ \d s ~,   \label{Y}
\ee
where the bracket $\{\,,\}$ is the Schwarzian derivative
\be
 \{f(z), z\} &\equiv& \frac{f'''(z)}{f'(z)} - \frac{3}{2}
 \left(\frac{f''(z)}{f'(z)}\right)^2 ~. \label{schwarz}
\ee
The condition for this process indeed to be a martingale is given
in Proposition 5.3 of Ref.~\cite{LSW}
\be
\lambda &=&  \frac{(3\kappa-8)(\kappa-6)}{2\kappa} \label{ak} \\
\alpha  &=&  \frac{6-\kappa}{2\kappa} \label{lk} ~.
\ee

Recall that we already found that a conformal form $\omega_t$ of
weight $h$ was a martingale $\hat{H} \omega_t =0$ provided
(\ref{ck}) and (\ref{hk}) hold. These parameter values are
consistent with the above-found parameter values, provided
$\alpha=h$ and $\lambda=-c$. Note, however, that the range of the
diffusion parameter $\kappa \in (0,8/3)$ restricts the central
charge to negative values. This was already noticed in
\cite{LSW,FW1}.

\section{Partition functions}
\label{partition}

\begin{figure}[ht]
\begin{center}
\includegraphics[scale=0.5]{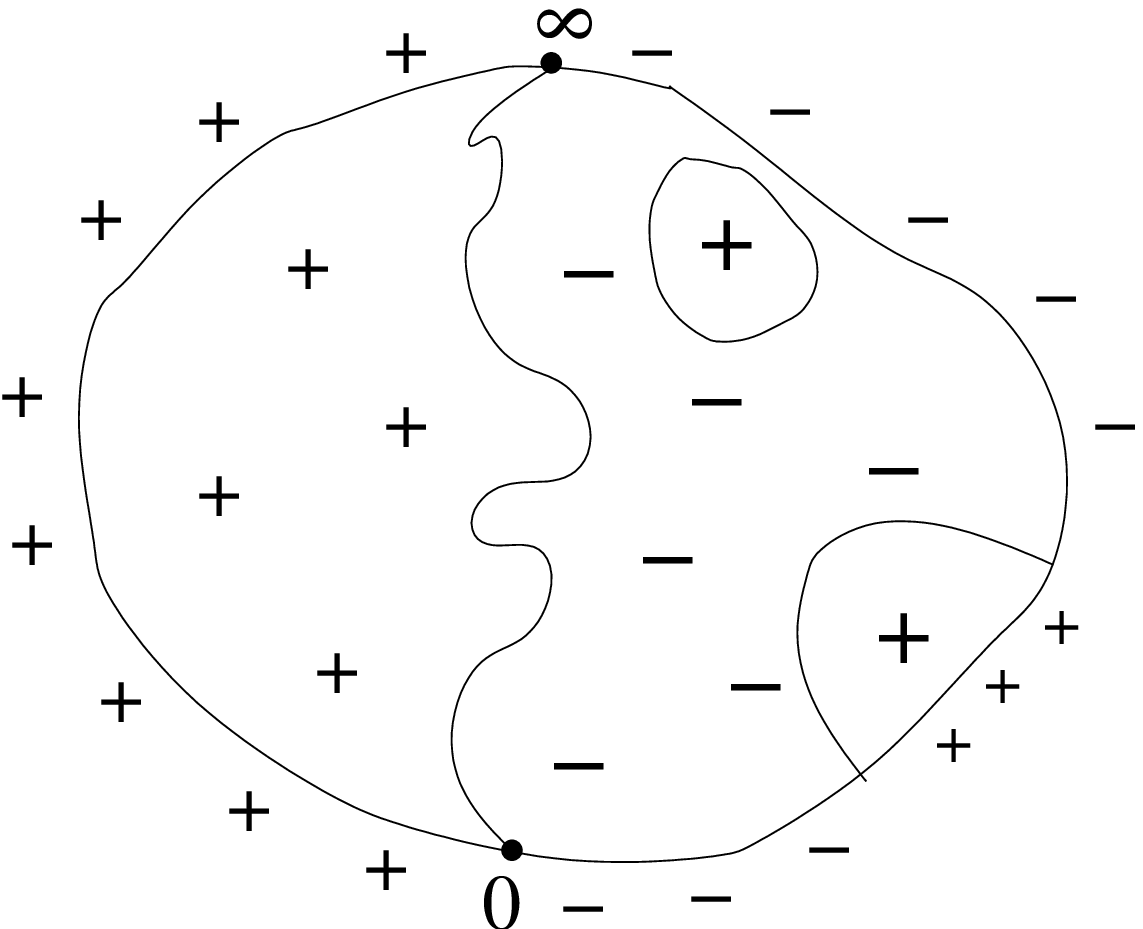}
\mycaption{A typical phase boundary $\gamma$ in a $\Zset_2$-model
on a simply connected domain $D$.} \label{phaseboundaries}
\end{center}
\end{figure}

Consider a statistical mechanics model on a discretisation of a
surface $X$ with states getting values in the discrete set  $\Zset_2$. At a 
critical temperature $T=T_c$ the correlation length diverges, and
the correlation functions become conformal. The typical statistical mechanics 
configuration below the critical temperature 
is patchwise constant, and the emerging structure is that of 
Kadanoff's droplet picture. The fact that the correlation length diverges 
at criticality means that the structure becomes self-similar. 
One can nevertheless still recognise a fractal phase boundary structure 
$\gamma \subset X$ as depicted in the heavily simplified 
Figure \ref{phaseboundaries}.

If we now fix boundary conditions we force the phase boundaries to
include in particular a path that connects the points on the
boundary where boundary conditions are changed. In  Figure
\ref{phaseboundaries}, for instance, this happens at $0$ and
$\infty$. Since the partition function with free boundary conditions, $Z$, should  provide a measure for the
number of states in the physical system, the partition function
$Z_{\alpha\beta}$ with fixed boundary conditions $\alpha\beta$
should provide a measure for the paths forced by changes of boundary conditions. Consequently, the fraction
\be
\frac{Z_{\alpha\beta}}{Z}
\ee
should be  the fraction of phase boundaries that include the paths forced
by these changes of boundary conditions among all possible phase
boundaries included in the full partition function. It can be
heuristically considered, therefore, to be the probability that
some path connects these specified boundary points in the
unconstrained theory.

The question of how to define properly
statistical mechanics models on Riemann surfaces is still, however, an open
problem. For present purposes we translate the problem to a
question in CFT, where this issue has been addressed in the Literature already:

At the critical temperature $T=T_c$, namely, the above statistical
mechanics models become scale invariant. We assume here that there
should then exist a CFT that has the same correlators as the
statistical model. For a definition of a CFT, see for instance 
Segal's axioms in \cite{Segal:sk,Charpentier:1990gn}. 
The partition function 
$Z \equiv\langle \unit \rangle$ can clearly still be thought of as a  weighted sum over configurations as in the statistical mechanics case; the same is true for the partition functions with different boundary conditions. Suppose we require the boundary conditions $|\tilde\alpha_i\rangle$ to apply on intervals labelled by $i=1,\ldots,n$. This can be done technically by inserting the
corresponding boundary operators $\phi_i \equiv
\phi_{\alpha_i\alpha_{i+1}}$ that change the state
$|\tilde\alpha_i\rangle$ to the state $|\tilde\alpha_{i+1}\rangle$
at the endpoints of the intervals $\{ x_1,\ldots,x_n\}$
\cite{Cardy:ir}, for general discussions see
\eg~Refs.\cite{DiFrancesco:nk,IR}. This can be used to express the 
partition function of the constrained theory in terms of the BCFT correlation function
\be
Z_{\alpha_1\ldots\alpha_n} &\equiv& Z \cdot  \langle \prod_{i=1}^n
\phi_i(x_i) \rangle ~.
\ee 
On physical grounds it seems reasonable to expect that the (positive) ratio 
$Z_{\alpha_1\ldots\alpha_n}/Z$ should be bounded by $1$; this issue should nevertheless be settled by carefully studying candidate CFTs case by case. We do not need to assume that there were only one boundary component. In what follows, we shall refer to these operator insertions collectively as $\Xope(x_i) \equiv \prod_{i=1}^n \phi_i(x_i)$.

In CFT we cannot restrict configurations to phase boundaries as in
the case of classical statistical mechanics models, simply because
configurations in a quantum field theory do not lend themselves to
classical treatment. The only well-defined restrictions are indeed
boundary conditions as discussed above.

Suppose we have inserted a boundary operator at
a marked point $\langle \Xope(0) \rangle_X$ and wish nevertheless
to restrict to states that have a quantum mechanical analogue of a 
phase boundary along some path $\gamma$ that starts from the
marked point and ends somewhere in the bulk $z \in X$. This is
actually possible if we first cut out the path from the domain
where the CFT is defined, thus in effect turning the path into a
part of the boundary where we can impose boundary conditions
$\langle \Xope(0) \rangle_{X \setminus \gamma}$, and then move the 
operator insertion to the tip of the cut-out path. The ratio
\be
\frac{\langle \Xope(p) \rangle_{X \setminus \gamma}}{\langle
\Xope(0) \rangle_{X}}
\ee
can therefore be considered as the probability associated to the
path $\gamma$ occurring among all the possible phase boundaries
forced by the operator insertion $\langle \Xope(0) \rangle_X$  at
the original point on the boundary.
We will next find this object for curves that correspond to
Loewner processes in (\ref{varE}) and (\ref{varEE}).

\section{Moduli under Loewner process}
\label{moduli}

A Riemann surface $X$, \ie~the complex structure of the two-dimensional manifold $X$, corresponds to a conformal class of metrics, which locally can be expressed in the form 
$\d s^2 = \e^\sigma|\d z|^2$, where $z$ is a local isothermal  coordinate. In general, given a two dimensional manifold with such a metric we can deform it in the following way: 

Deformations of a metric $g$ on the Riemann surface $X$ can be
decomposed to local reparametrisations given by a global vector
field $v \in \Gamma(TX)$ on the surface, local Weyl rescalings given by a global function $\varphi$ on $X$, and Teichm\"uller
reparametrisations given by the Beltrami differentials 
\be
\mu \in
\Omega^{(-1,1)}(X) \equiv T^{(1,0)}X \otimes \Omega^{(0,1)}(X) 
\ee
and $\bar\mu\in \Omega^{(1,-1)}(X)$. Under these transformations a
correlator
\be
\langle \Xope \rangle \equiv \langle \prod_i \Phi_i(z_i) \rangle
\label{xope}
\ee
of (holomorphic) primary fields -- of given spins $s_{i}$ and scaling dimensions $\Delta_i = 2 h_i - s_i$ -- inserted at points $z_i \in X$ transforms as \cite{DiFrancesco:nk}
\be 
\langle \delta \Xope \rangle 
&=& \frac{1}{2\pi i} \int_X
\d^2z \Big[  (\nabla_{\bar{z}} v^{z} + \mu) \langle T(z) \Xope
\rangle + (\nabla_{{z}} v^{\bar{z}} + \bar\mu) \langle
\bar{T}(\bar{z}) \Xope \rangle  \Big] \nonumber \\ 
&& - \sum_i
\Delta_i \varphi(z_i) \langle   \Xope \rangle \label{varX}~. 
\ee 
The last term here arises from delta-functions due to insertions of $T_{z\bar{z}}+T_{\bar{z}z}$, though only when we choose to change the representative of the conformal class $\sigma \longrightarrow \sigma + \varphi$ as well. We use here conformally flat reference metric proportional to $|\d z|^{2}$. Choosing the transformations judiciously, this identity implies also the standard conformal Ward identity.

On manifolds with boundary, the diffeomorphisms generated by
the vector field $(v^z, v^{\bar{z}})$ are required to preserve the
boundary; this necessitates also that only one independent copy of
$\Vir$ and $\overline{\Vir}$ is preserved so that at the boundary
$x \in \partial X$ the stress-energy tensors coincide $T(x) = \bar{T}(x)$.

As the above deformations are all of the deformations we can
perform in two dimensions (conformal and complex structures being
equivalent), then the Beltrami differentials are the only true
deformations of the moduli of the theory, and can be thought of as
(anti-)holomorphic vector fields on the tangent space of the
moduli space $(\mu, \bar{\mu}) \in  T_X{\cal M}$. More formally,
the Beltrami differentials $\mu$ can be thought of as classes of
the tangent bundle valued first cohomology group $\mu, \bar{\mu}
\in H^1(X,TX)$. In the case of the moduli space of Riemann
surfaces with boundary the moduli space has, again, only real
analytic structure.

The partition function $Z$ depends, first of all, on the moduli of
the Riemann surface $X$ and the details of the CFT defined on that
surface \cite{Friedan:1986ua}. It is therefore, in particular,
locally a function of the coordinates $m,\bar{m}$ of the moduli
space. In a nontrivial CFT $c\neq 0$ there is a trace anomaly
and the partition function depends on the choice of a representative
of the conformal class. In defining the partition function in this
way we need to specify, therefore, that the partition function
$Z(m,\bar{m})$ is evaluated, for instance, in the constant
curvature background metric $g_\cc$ in the conformal equivalence
class of metrics we are interested in. The dependence on the
representative of the equivalence class arises through the
Liouville action $S_L(\sigma,g_\cc)$ so that if we know the
partition function in the constant curvature background metric
$Z(m,\bar{m})$, on general backgrounds $g = \e^{\sigma} g_\cc$ the
physical partition function becomes
\be
Z(g) &=& \e^{{c}S_L(\sigma,g_\cc)} ~Z(m,\bar{m}) ~. \label{weyl}
\ee

Neither is the partition function $Z(m,\bar{m})$ 
in general a well-defined
function on the moduli space of Riemann surfaces, but rather a
section of a (projective) line bundle on it. 
In the case of closed surfaces this line bundle factorises $E_c
\otimes \bar{E}_c$ to holomorphic and anti-holomorphic parts. The
line bundle $E_c$ comes equipped with a 
connection, with respect to which the partition function is
covariantly constant. This can be
seen also form Eq.~(\ref{varX}) by choosing $\Xope=\unit$ so that
$\delta \Xope = \delta \unit = 0$. This implies
\be
\nabla_\mu  Z \equiv
\delta_\mu Z - \frac{1}{2\pi i} \int_X \d^2 z ~ T(z)
\mu(z,\bar{z}) ~ Z = 0 ~, \label{covaT}
\ee
which tells us that the partition function is parallel transported
along the vector field $\mu$ on the moduli space with respect to the 
connection $\d + T$.
The holomorphic part can be recognised as a
tensor power of the standard determinant bundle 
\be 
E_c &=&
{\det}_X^{\otimes c/2} 
\ee 
otherwise known as the inverse Hodge bundle $\Det_1 = {\det}_X^{-1}$.

In the case of Riemann surfaces with boundary components the
holomorphic and the anti-holomorphic sectors are related by
complex conjugation. This means that the partition function 
$Z(m,\bar{m})$ with $m=\bar{m}^*$  is
actually a section of the emerging real-analytic 
bundle 
\be
{\Xdet}^{\otimes c} \longrightarrow {\cal M}~.
\ee
Boundary operators can be described\footnote{At least in the unitary case.}  in terms of bulk operators on the Schottky double by insertions of operators and their complex conjugates on the original surface, and its mirror image, respectively. This means in particular that if in the expectation values of bulk operators inserted at the point $z \in X$ transform as elements of $(T^{*(1,0)}_{z}X)^{\otimes h'}$ for some conformal weights $h'$, then the corresponding \cite{DiFrancesco:nk} expectation values of  boundary operators at $z \longrightarrow x \in \partial X$ transform as elements of $|T^{*}_{x}\partial X|^{\otimes h}$ for some conformal weights $h$ that depend on the structure of the theory at hand.

\begin{figure}[ht]
\begin{center}
\includegraphics[scale=0.6]{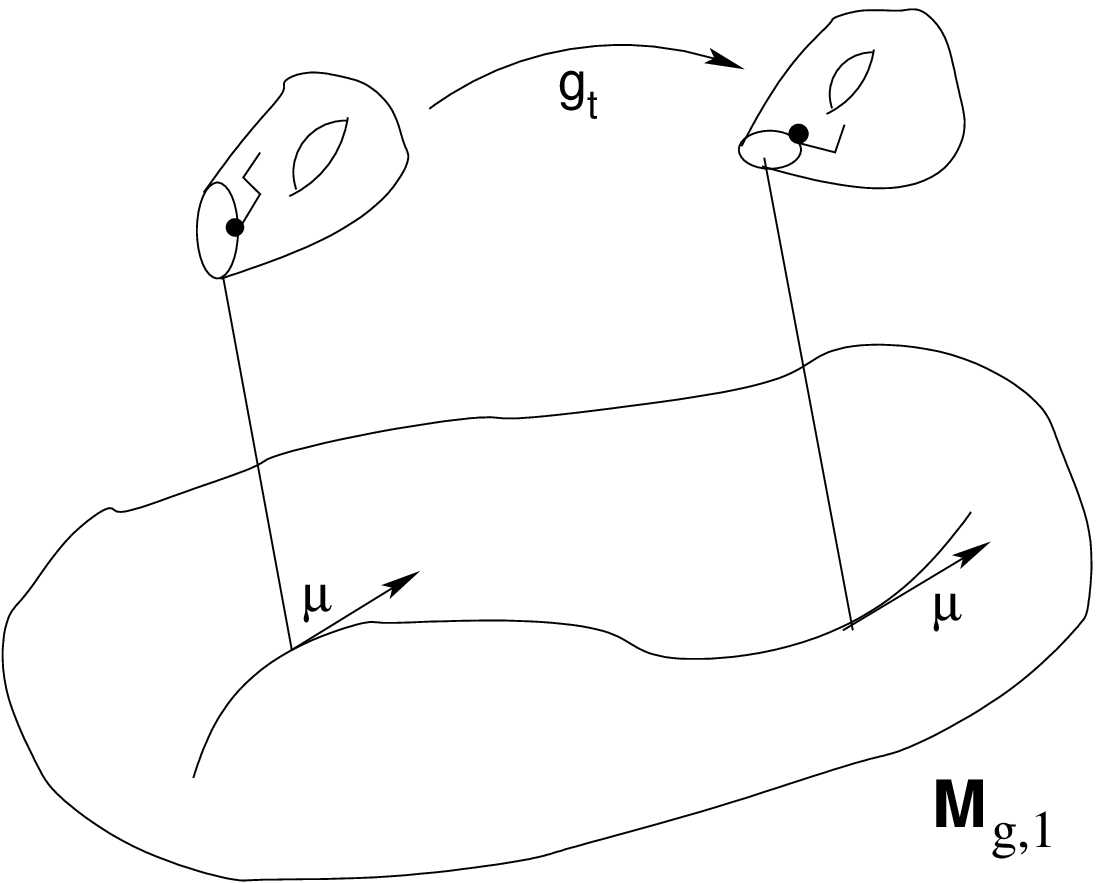}
\mycaption{Changing the Riemann surface in Loewner process $g_t$
induces a random walk in the moduli space with the tangent vector field
given by the Beltrami differentials $\mu_t$.} \label{modspace}
\end{center}
\end{figure}


\subsection{Infinitesimal deformations} 
\label{newSec}

Eq.~(\ref{covaT}) above is an example of the natural pairing given by integration over the Riemann surface $X$ of Beltrami differentials $\mu \in  T_X{^{(1,0)}\cal M}$ and holomorphic quadratic differentials $\nu \in \Omega_X^{(2,0)}(X)$ 
\be
(\nu,\mu) &\equiv& \int_X \nu \wedge \mu ~, 
\ee 
as the stress energy tensor $\nu = T(z) (\d z)^2 \in \Omega^{(2,0)}(X)$ 
is a locally defined quadratic differential. Holomorphicity is required here for guaranteeing independence of the choice of the representative of the Beltrami differential $\mu \sim \mu + \bar\partial v$. On manifolds with boundary we must restrict to vector fields $v$ that generate flows that leave the boundary invariant. In this sense we can identify $\Omega_X^{(1,0)}{\cal M} \simeq \Omega^{(2,0)}(X)$.  

Let us consider Loewner process along a parameterised path $\gamma \subset X$, and suppose that we have divided it in infinitesimal parts $\gamma = \bigcup_{i} \gamma_{i}$. By infinitesimal we simply mean that the pertinent measure on each path $\gamma_{i}$ -- Hausdorff or Lebesgue --  is arbitrarily small. In what follows we denote this measure by $\d t$, given a specific parameterisation of the original curve.  

Consider now in particular Loewner procedure over the first of these infinitesimal instalments $\gamma_{0}$; the idea is to  iterate the procedure over all of the infinitesimal contributions to get a finite result, as we shall see: 
The Beltrami differential associated to cutting the surface along the  infinitesimal path $\gamma_{0}$ can be thought of as an infinitesimal vector $\mu_t \in T_X {\cal M}$. Its length is proportional to the volume of the cut-out set, \ie~$\d t$. It has support only on the path $\gamma_0$, if anywhere. Given, as a test function, a cotangent vector from $T_X^*{\cal M}$ represented by a smooth quadratic differential $\nu \in \Omega^{(2,0)}(X)$ the  pairing is, therefore, of the form 
\be
\Big(\mu_t,\nu(z) (\d z)^2 \Big) \sim \d t ~ \nu(\gamma_0)~.
\ee
Here $\nu(z)$ is assumed constant across the infinitesimal set $\gamma_0$. This means that the Beltrami differential can be put formally in the form
\be
\mu_t &\equiv& \mu_t(z) \frac{\d \bar{z}}{\d z} = - 2\pi i ~ \delta({\gamma_0}) \frac{\d t}{(\d {z})^2} ~,   \label{formalMu}
\ee 
where the formal current 
\be
- 2\pi i ~ \delta({\gamma_0}) {\d t} &=& \mu_t(z) \d^{{2}} {z} \label{msures}
\ee
has distributional support on the cut-out path $\gamma_0$. The details of the distribution will depend of how we choose to regularise it. 

Under Loewner procedure the conformal class of the original Riemann surface $X$ changes as a function of the parameter $t$ along the path $\gamma \subset X$. The resulting family of Riemann surfaces $X_{t} \subset {\cal M}$ traces similarly over a certain path $\Gamma \subset {\cal M}$ in the moduli space. The choice of  regularisation of the formal current above amounts then to a choice of parameterisation of this curve as we shall presently see. 

To make the discussion indeed somewhat more concrete, let us return to the Eq.~(\ref{lemBeltrami}) where the Beltrami differential associated to the specific Loewner process $g_t$ on the upper half-plane was found to be   
\be
\mu_t(z,\bar{z}) &=& -2\pi i ~ \frac{\alpha}{1-\alpha} (z-t)^2 \delta^{(2)}(z-t) ~. 
\ee 
The distribution $\delta({\gamma_0})$ in this case is 
\be
\delta({\gamma_0}) &=& \frac{\alpha}{1-\alpha} (z-t)^2 \delta^{(2)}(z-t)  ~. 
\ee
In view of the regularisation procedure that will follow, we choose first to change the parameterisation of the path $\Gamma$ from $t $ to $t'$ where
\be
\frac{\alpha}{1-\alpha} &=& \frac{\d t'}{\d t}~.  
\ee 
This is precisely the definition of the parameter on the path $\Gamma \subset {\cal M}$ alluded to below Eq.~(\ref{formalMu}).

This distribution is zero when integrated with regular test functions; in our case it will appear with functions with second order poles precisely at $z=t$ 
\be
\langle T(z) \phi_h(t,\bar{t}) \rangle &\sim& \frac{h}{(z-t)^2}  
\phi_h(z,\bar{z}) + \frac{1}{z-t} \partial_z \phi_h(z,\bar{z}) ~,
\ee 
so that it picks the coefficient of the leading pole in the operator product expansion, provided the operator $\phi_h$ is inserted precisely at $z=t$. It is therefore useful to define the operator $\hat{T}$ that does just that, namely picks the leading term in the Laurent expansion of the stress energy tensor with operator insertions at the specified point, for instance
\be
\hat{T}(z) \cdot \phi_{h}(t)  &:=& h~ \phi_h(t,\bar{t}) ~,
\ee 
for $z=t$, otherwise $\hat{T}(z) \cdot \phi_{h}(t) =0$.  
One could represent this operator \eg~in terms of contour integration of the standard operator product. 
All other operator insertions provide a trivial result. Note also that the contribution coming from the operator insertion at $z=t$ will now be finite, and its precise value does indeed depend on the regularisation or other details of the distribution $\delta({\gamma_0})$.  

We may now express the change of the correlator through 
\be
\Big\langle (\mu_{t}, ~ {T} ) \cdot \phi_{h}(t) \Big\rangle_{\text{regularised}} &=&  \langle 
\d t' ~ \hat{T}(t) \cdot \phi_{h}(t) \rangle
\ee
which is now to be looked upon as a one-form in  $T^*_X {\cal M}$. Note that the argument of $\phi_{h}$ is a point on the boundary of the Riemann surface and the differential $\d t'$ refers to a parameterisation of the path $\Gamma \subset {\cal M}$.

The operator insertions we consider $\phi_{h}$ are precisely where we want the boundary conditions in CFT to change -- at the intersection of the considered path $\gamma_{t}$ and the boundary $\partial X$. After having performed the above iterative step, we should, therefore, translate the operator insertion from the original point $z=t$ to $z=0$ where the rest of the path  intersects the boundary. This is to be seen as a part of what we mean by Loewner procedure, but boils down to choosing to evaluate the field $\phi_{h}(z,\bar z)$ precisely at $z=0$; the difference is indeed negligible as long as the infinitesimal path $\gamma_{0}$ is small enough. This is also reminiscent of the definition of stochastic integrals, where one chooses to evaluate the integrand on the left point of each interval.  
 
Inserting these results in  Eq.~(\ref{varX}) and not forgetting the anti-holomorphic sector produces now 
\be
\langle \delta \Xope \rangle 
&=& -  \d t' ~ \Big[ \langle \hat{T}(\gamma_0(t)) \Xope \rangle + \langle \hat{\bar{T}}(\gamma_0(t)) \Xope \rangle \Big] ~,
    \label{varB}
\ee 
with the understanding that the right-hand side involves an insertion of $\Xope$ in the beginning of the path and the left-hand side in the end of the path. 
In summary, the concrete analysis has produced the following results: 
\begin{itemize}
\item[1)]
With an operator insertion at the intersection of the path $\gamma$ and the boundary $\partial X$, correlators transform 
by an infinitesimal but nontrivial amount;  
\item[2)] 
The transformation can be expressed in terms of a (finite) line integral of the stress-energy tensor along the path $\gamma \subset X$; and, 
\item[3)] 
The integration parameter is determined in regularising Eq.~(\ref{msures}); the integral is invariant under reparameterisations as long as we change the volume measure on the Riemann surface $X$ on the right-hand side of this equation or the regularisation of the distribution on the left-hand side as well.  
\end{itemize}

Let us now return to the general discussion, assuming we have regularised the distribution $\delta(\gamma_{0})$, introduced operators $\hat{T}$, and a parameterisation $t$ of the path $\Gamma \subset {\cal M}$: 

Iterating this procedure for all infinitesimal paths $\gamma_{i}$ 
is tantamount to exponentiating the infinitesimal operation: the procedure leads indeed to (essentially) the standard path-ordered exponential that appears in parallel transportations, which  can in this case be defined in a regularised form, in notation that will be explained below,  as  
\be
\Pexp  - \int_{\gamma} {T}(\gamma_t) \d t  &:=& \prod_{i} \Tset( \gamma_i) \Big( \unit  - \int_{\gamma_{i}}\d t  ~ \hat{T}(p_{i}^{0})   \Big)  ~. \label{regu}
\ee  
Here the stress energy tensor is evaluated in the starting point $p_{i}^{0} \in \partial \gamma_{i}$ of each infinitesimal path $\gamma_i$, and the integral itself reduces to the (Lebesgue or Hausdorff) measure of the infinitesimal path $\int_{\gamma_{i}} \d t$. Now, in the limit where the paths are indeed taken arbitrarily small, the operator product expansion of the factors in this formula with any operator insertion $\phi_{h}(x)$ on the boundary $x \in \partial X$ are never more singular than with insertions at the beginning of each path $p_{i}^{0} \in \partial\gamma_{i}$ for the simple reason that this is the only place where the infinitesimal path intersects the boundary: Therefore, all operator product expansions are dominated by whatever contribution arises from the starting points $x=p_{i}^{0}$ and, as was shown above, these contributions are finite. We have stipulated a specific way of doing this expansion by determining that the operator insertion $\phi_{h}$, if any, be translated  from the beginning of the path $p_{i}^{0}$ to the end $p_{i}^{1}$
by inserting explicitly the translation operators
\be
\Tset(\gamma_{i}) \cdot \phi_{h}(p_{i}^{0}) & := & 
\phi_{h}(p_{i}^{1})  ~. 
\ee 

What all of this amounts to is a specific regularisation of the 
formal, a priori perhaps rather singular operator 
\be
\Pexp  \Big( - \int_{\gamma} {T}(\gamma_t) \d t \Big) \cdot \phi_{h}  ~. 
\ee 
There might be other ways of regularising this object -- essentially a parallel transport of operators along the path $\gamma$. It would be interesting to investigate further how different regularisations affect the present discussion or under what conditions this formal product should indeed converge. In particular, one should consider insertions of boundary operators in terms of insertions of bulk operators together with their mirror image on the other half of the Schottky double. 
We leave these issues, however, to later study.


\subsection{Parallel transport}

Let us consider now the correlation function
\be
{\langle \Xope(y_i) \rangle}_{\gamma[0,t]} &\equiv&  \Big\langle \Pexp -
\int_{\gamma[0,t]} \d s \Big( T(\gamma(s))
     + \bar{T}(\gamma(s))\Big) \cdot \Xope(y_i)  \Big\rangle
     \label{varE} ~,
\ee
associated to the path $\gamma[0,t]$ that follows the path $\gamma$ from time $0$ until the time $t$. The definition of the path-ordered exponential used here was the subject of Sec.~\ref{newSec}; the regularised expression we intend to use here was explicitly constructed in Eq.~(\ref{regu}). In particular, the exponential is to be expanded  
in a product of exponentials of infinitesimal contributions 
along the path acting successively on $\Xope(y_i)$, and the operator insertion should be transported along the path in the process of performing the integration. The operator $\Xope$ was defined in Eq.~(\ref{xope}) in terms of chiral primaries. In what follows we shall choose to restrict to operator insertions on the boundary $z_i = y_i \in \partial X$, and choose the operators themselves from the pertinent BCFT. These operators are, by construction, invariant under complex conjugation and the associated correlators are thus real. As was observed in the end of Sec.~\ref{moduli}, they transform as elements of $|T^{*}_{y_{i}}\partial X|^{\otimes h_{i}}$ under conformal transformations. 

Using the explicit Beltrami differential (\ref{lemBeltrami}), it
was possible in fact to show in Sec.~\ref{newSec} that the Loewner procedure along the path $\gamma$ maps the correlator ${\langle \Xope \rangle}$ to the above correlator ${\langle \Xope \rangle}_{\gamma[0,t]}$. What we see here is, therefore, its parallel transport with respect to the operator valued connection $T$ \cite{Ranganathan:1993vj}. Physically one can think of this as the partition function in the presence of energy density, or a current, distributed along the path $\gamma$.

The argument to this effect 
made use of the explicit form of the
Beltrami differential $\mu$ on $\Hset$ given in Eq.~(\ref{lemBeltrami}),
and the fact that it 
always arises together with the correlator of the
stress-energy tensor $T$ 
and the explicit operator insertion $\Xope$: The 
operator product expansion has precisely the quadratic
pole needed to produce a nontrivial result. 
The infinitesimal change in the correlator can be thought of as a
suitably normalised differential on the moduli space; Cutting out parts of
the path repeatedly leads to the path-ordered exponential integrated along
a finite path in the moduli space. 

We can now define a probability density associated to any parametrised path in the space of paths on the surface $X$ 
$\Pi(X,p;t)$, namely let 
\be
{\cal P}_X : \Pi(X,p;t) \longrightarrow \Rset^*_+ ~ ; ~ \gamma
\mapsto  \frac{{\langle \Xope \rangle}_{\gamma[0,t]}}{{\langle \Xope
\rangle}} \label{main}
\ee
for each path in $\Pi(X,p;t)$ that starts from a fixed point 
$p \in \partial X$ and goes on until the final parameter value 
$t$. This density has, first of all, the property that it
is real and normalised ${\cal P}_X \in (0,1]$ such that ${\cal
P}_X(p) = 1$. Reality follows from the facts that the operator
insertions on the boundary are, by construction, real $\Xope^* =
\Xope$ and that the conformal mapping $f : X\setminus \gamma
\longrightarrow X$ preserves the boundary on the real axis. The
fact that it is non-negative follows from the assumption that
$\langle \Xope \rangle \neq 0$ as a physical 
partition function, and the fact that $\langle \Xope \rangle_{\gamma[0,t]}$ 
was constructed from it by exponentiation.  It is bounded by its value ${\cal
P}_X(p) = 1$ if we assume that the classical weak energy condition
 ${T}+{\bar{T}}>0$ translates to a positivity 
condition on the spectrum of the corresponding operator in CFT. 
Whether these (mild) assumptions are satisfies 
depends on the details of the pertinent CFT model.

The above observations amount to the statement that we can
consider ${\cal P}_X(\gamma)$ to be a probability associated to a
path $\gamma$. It is, however, {\em not} the probability of the
path occurring among all the paths of $\Pi(X,p)$ 
but rather the probability of finding a path in a
hull of $\gamma$, perhaps, whose width is related to the structure
of the CFT, and the central charge \ie~the diffusion coefficient
in particular. The precise interpretation of this probability
density follows from stochastic considerations, and will be
deferred to Sec.~\ref{disc}.

Suppose now that the simple curve $\gamma[0,t]$ is an
SLE${}_\kappa$ process $\gamma_t$ on a Riemann surface $X$ 
with $0 \leq \kappa < 4$. We can then likewise associate to this path
the correlator
\be {\langle \Xope \rangle}_{\gamma[0,t]} &=&  \Big\langle \Pexp -
\int_{\gamma[0,t]} \d \sigma_H(s) \Big( T(\gamma(s))
     + \bar{T}(\gamma(s))\Big) \Xope \Big\rangle
     \label{varEE} ~,
\ee
where $\sigma_H$ is the Hausdorff measure of dimension \cite{B}
\be
\dim_H(\gamma) &=& \min \left(2, 1 + \frac{\kappa}{8} \right)~.
\label{Bdim}
\ee
This extends the definition (\ref{varE}) to the case of fractal
curves. It shows, again, how the correlation functions of 
observables ${\langle \Xope \rangle}$ transform under Loewner
processes, and it continues to be normalised as suggested above.

\subsection{Conditioning correlators}

A local conformal transformation $\rho$ induces a transformation
$R(\rho)$ of the pertinent CFT (Hilbert) space.\footnote{This amounts to a representation of $\Aut~{\cal O}$ defined in Eq.~(\ref{auto}) on the pertinent Hilbert space constructed essentially  by exponentiating the positive part of Virasoro algebra $\Vir_{\geq 0}$, \cf~\eg~Lemma 5.2.2 in Ref.~\cite{FB}.} The operator insertions of primary fields transform homogeneously, whereas the stress-energy operator changes inhomogeneously
\be
\Xope (y_i) &=& R(\rho) \Xope \Big(\rho(y_i)\Big)R(\rho)^{-1} ~
\prod_i \Big( \rho'(y_i) \Big)^{h_i} \\
T(z) &=& R(\rho)  T\Big(\rho(z)\Big)  R(\rho)^{-1} + \frac{c}{12}
\Big\{ \rho(z), z \Big\} ~ \unit
\ee
where $\{ ~ ,~ \}$ is the Schwarzian derivative (\ref{schwarz}).
The correlator ${\langle \Xope \rangle}_{\gamma[0,t]}$ transforms
therefore as
\be
\rho^*{\langle \Xope  (y_i) \rangle}_{\gamma[0,t]} 
&=& \prod_i \Big( \rho'(y_i)
\Big)^{h_i} ~ \exp -\frac{c}{6} \int_{\gamma[0,t]} 
\d s ~ \Re\Big\{ \rho(\gamma(s)), \gamma(s) \Big\} \nonumber \\
  & & \qquad  \qquad  \cdot {\langle \Xope  (y_i) 
\rangle}_{\gamma[0,t]} \label{transfR} ~. 
\ee

In order to condition the probability to paths that do not enter 
a given simply connected domain $A \subset X$ that touches the
boundary, we need to investigate the behaviour of the density 
${\cal P}_X$ under the pertinent diffeomorphism $\rho : X \setminus A
\longrightarrow X$. In the simple case that this diffeomorphism happens 
to be a conformal mapping or that the pertinent moduli space is a discrete
set of points as in the case of the upper half-plane $\Hset$ we find 
\be
\frac{ {\cal P}_{X \setminus A} }{ {\cal P}_X } &=& \prod_i \Big(
\rho'(y_i) \Big)^{h_i} ~ \exp -\frac{c}{6} 
\int_{\gamma[0,t]} \d s ~
\Re \Big\{ \rho(\gamma(s)), \gamma(s) \Big\} ~.
\ee

Comparing to the process $Y_t$ in the stochastic analysis
(\ref{Y}), we see that ${\cal P}_X$ and $P$ have precisely the
same behaviour under conditioning, \cf~Sec.~\ref{disc}. In
comparing this expression to $Y_t$ in stochastic analysis, one
needs to take into account the fact that there it was convenient
to keep the origin of the complex plane fixed and let the
intersection $W_t = \gamma_t \cap
\partial \Hset$ move, whereas in the CFT analysis one kept the
intersection fixed at the origin and allowed the original zero to
move.

In the general case where the diffeomorphism  $\rho : X \setminus A
\longrightarrow X$ changes the conformal structure of the Riemann surface, 
it is difficult to give an explicit formula for the transformation. 
For infinitesimal such deplacements this is nevertheless possible, and 
reduces clearly to insertions of the stress-energy tensor 
integrated with the Beltrami 
differential associated to $\rho$ in the correlators.

The correlator $\langle \Xope \rangle_{\gamma[0,t]}$ can be recognised as
a section of a certain bundle $\Obdle_h$ over
the moduli space of Riemann surfaces.\footnote{For the  definition, see Eq.~(\ref{obdleDef}).} In this context it is clear
that ${\cal P}_X(\gamma)$ is consequently a holonomy, or a Wilson
line, of this section when parallel transported from the fibre at
$X$ to the fibre over ${X \setminus \gamma}$ with respect to the
connection  $\d + T$. 

Recall that if the correlator $\langle \Xope \rangle$ satisfies 
$\hat{H} \langle \Xope \rangle =0$, the operator creates a state in the 
Verma module $V_{2,1}$. This module is closed under Virasoro action, 
which in turn is generated by the stress-energy tensor $T$. Since 
the Loewner process involves only insertions of the stress-energy 
tensor in the correlator, the final correlator  
$\langle \Xope \rangle_{\gamma[0,t]}$ has to be that of an operator 
belonging to the same Verma module
and satisfying the same differential equation. This is true irrespective
of the moduli of the Riemann surface, and provides indeed an independent 
analytic characterisation of the correlators 
$\langle \Xope \rangle_{\gamma[0,t]}$ as those sections of  
$\Obdle_h$ that are annihilated by $\hat{H}$. 

In the next section we will describe this bundle 
$\Obdle_h$ and the action of $\hat{H}$ in detail.

\section{Determinant bundles}
\label{det}

In this section we shall consider the geometry of the densities
${\cal P}_X$ as sections of certain bundles over moduli space. We
shall pay particular attention to the explicit form of the
Virasoro action on these bundles because it translates directly to
a description of the Loewner process in the moduli space. We follow in much the notation of Ref.~\cite{FB}.

\subsection{Virasoro action}
\label{vira}

The CFT partition functions on Riemann surfaces with $b$ boundary
components and of genus $g$ could be seen as a section of the
determinant bundle
\be
\langle \unit \rangle &\in& \Gamma\Big(  {\cal M}_{g,b}, 
{\Xdet}^{\otimes c} \Big)
\ee
as discussed in Sec~.\ref{moduli}.  
We recall that the fibre at
$X \in {\cal M}_g$ of the standard determinant bundle $\Det_j$
with $j \in \Zset$ is
\be
\bigwedge^{\max}  H^0(X, \Omega_X^{\otimes j}) \otimes
\bigwedge^{\max} \Big( H^1(X, \Omega_X^{\otimes j}) \Big)^*~.
\ee
The determinant bundle ${\det}_X$ associated to a surface $X$ is
the inverse of the Hodge bundle ${\det}_X = \Det_1^{-1}$. Of
determinant bundles it is known, for instance, that the Hodge
bundle $\Det_1$ generates \cite{AC}  the Picard group of ${\cal
M}_g$, and that $\Det_j \simeq \Det_1^{\otimes(6j^2-6j+1)}$
\cite{Mum}.

For one (spin-0) operator insertion $k=1$ at $y_1=p$ with conformal weight $h$ the transformation rule (\ref{transfR}) implies ${\langle
\Xope \rangle}_{\gamma[0,t]} \in \Gamma({\Obdle}_h)$ where\footnote{To be quite concrete, a point in the moduli space $(X,p) \in {\cal M}_{g,1}$ determines a Riemann surface and a marked point $p \in X$. As the fibre in  ${\Xpdet}^{\otimes c}$ consists of nonnegative numbers $v$, we can give a point in the total space as 
$(v ; X,p) \in {\Xpdet}^{\otimes c}$. The transformation functions $\lambda_{ij}'$ in this bundle can be given in terms of the transformation functions $\lambda_{ij}$ (\cf~Eq.~(\ref{weyl})) on $\det_{X,p}$ by setting  
$\lambda_{ij}' = |\lambda_{ij}|^{c}$. Twisting by $|T_p^*  \partial  X|^{\otimes h}$ means that if the representative of the conformal class of $(X,p)$ are different on two charts -- \ie~ we must perform a conformal transformation $\rho$ on the Riemann surfaces when changing charts in the moduli space -- then the full transition function in $ {\Obdle}_h$ is indeed   $|\lambda_{ij}|^{c} \cdot |\rho'(p)|^{h}$.}
\be
{\Obdle}_h &\equiv& {\Xpdet}^{\otimes c} \otimes |T_p^* \partial X|^{\otimes h} ~. \label{obdleDef}
\ee 
Here we have used the pull-back bundle 
$\Xpdet \equiv \varpi^*\Xdet$ where $\varpi : {\cal M}_{g,1} 
\longrightarrow {\cal M}_{g}$. The fibre of ${\Xpdet}^{\otimes c}
\longrightarrow {\cal M}_{g,1}$ at the Riemann surface $X$ is
twisted by a tensor power of the modulus of the cotangent space of the surface at the marked point $p \in \partial X$. The modulus appears, as was discussed in the end of Sec.~\ref{moduli}, because insertions of boundary operators can be described as insertions of bulk operators and their mirrors on the Schottky double; as the mirror transforms by the complex conjugate $\rho'(\bar p)^{*}$ of the transformation of bulk field\footnote{Note that the conformal weight $h$ above is the conformal weight of the associated boundary field, not that of the bulk field and its mirror, as discussed in the end of Sec.~\ref{moduli}.}
$\rho'(p)$, the total effect is a transformation by a positive function $|\rho'(p)|^{2}$.

Let us denote by ${\cal O} = \Cset[[z]]$  the set of formal Taylor
series of a formal coordinate on the unit disc $z \in \Dset$, and
${\cal K} = \Cset((z))$ the set of formal Laurent series in the
punctured disc $z \in \Dset^*$. The automorphism group $\Aut~{\cal
O}$ are changes of coordinates, representable in terms of Taylor
series of the form
\be
\Aut~{\cal O} \simeq \Big\{ \sum_{n=1}^\infty a_n z^n : a_1 \neq 0
\Big\}. \label{auto}  
\ee

We associate to the marked point $p$  a formal coordinate $z \in
\Dset$ that gets its values in the formal unit disc. The triples
$(X,p,z)$, where $X$ is the Riemann surface, glue together to an
$\Aut~ {\cal O}$-bundle
\be
\pi : \widehat{{\cal M}}_{g,1}
\longrightarrow {\cal M}_{g,1} ~.
\ee
The action of $\Aut ~ {\cal O}$ represents changes of formal
coordinates, and has the Lie algebra $\Lie ~\Aut~ {\cal O} =
\Vir_{\geq 0} = z \Cset[[z]] \partial_z$. We can think of the
fibres of this bundle, therefore, as the set of all choices of
formal coordinates around the marked point $p \in X$.
$\widehat{{\cal M}}_{g,1}$ can also be viewed as a bundle 
\be
\hat\pi : \widehat{{\cal M}}_{g,1} \longrightarrow {\cal M}_g ~,
\ee 
but we have to add to the structure group also the shifts of
the marked point, generated by $\partial_z$. Changes of the
conformal structure of the surface are generated at the fixed
point by the singular vector fields of the form $z^{-n} \Cset[[z]]
\partial_z \subset \Vir_{<-1}$ for $n>0$. All these formal vector
fields included in $\Vir_{<-1}$, $\Vir_{\geq 0}$, and $\Cset
\partial_z$ form together the Witt algebra $\Der ~{\cal K}$. The
actual Virasoro algebra $\Vir$ is a central extension of this
(\ref{cextension}).
One can actually show that $\widehat{{\cal M}}_{g,1}$ carries a
transitive action of $\Der ~{\cal K}$ compatible with the $\Aut~
{\cal O}$ action along the fibres \cite{K,BS}, \cf~also \cite{FB}.

The underlying structure here is the Harish--Chandra pair $(\Der
~{\cal K}, \Aut~ {\cal O})$ and a flat vector bundle
$\widehat{{\cal M}}_{g,1} \longrightarrow {\cal M}_{g,1}$
associated to it. This structure generalises \cite{K,BS} to the
Harish--Chandra pair $(\Vir, \Aut~ {\cal O})$ as well: it has an
associated flat vector bundle  $\hat\pi^*\Det_j \longrightarrow
\widehat{{\cal M}}_{g,1}$ whose infinitesimal automorphisms
$\Lie~\Aut~\hat\pi^*\Det_j$ form a representation of the Virasoro
algebra \cite{K}. The flat connection there is of the form $\d + L_{-1}$. 
Note that even if the tensor powers of these 
bundles were not globally well-defined, the underlying D-module is
always well-defined.

The difference between the bundles $\Obdle_h$ and
$\pi^*\Obdle_h$ is therefore that the latter extends the
former by keeping track of the formal coordinates near the marked
point. Where the curvature of (the holomorphic part of) 
$\Obdle_h$ was related to the Picard
group of the moduli space, the natural connection on the bundle
$\pi^*\Obdle_h$ is flat.

\subsection{Loewner process on the determinant bundle}
\label{det-loew}

As the Loewner process produces a nontrivial Beltrami
differential, it generates a motion in the pertinent moduli 
space ${\cal M}_{g,1}$. This involves 
deforming the surface, changing the complex structure, 
and displacing the marked point $p \in X$ on the surface. 
If we attach a formal disc with coordinate $z \in \Dset$, we 
see immediately that the above operations induce actions of 
$\Vir_{\geq 0}$, $\Vir_{<-1}$,and $\Vir_{-1} \sim \Cset \partial_z$
on the disc. This means that the Loewner process acts 
on the disc by  $\Der ~{\cal K}$. 

\begin{figure}[ht]
\begin{center}
\includegraphics[scale=0.5]{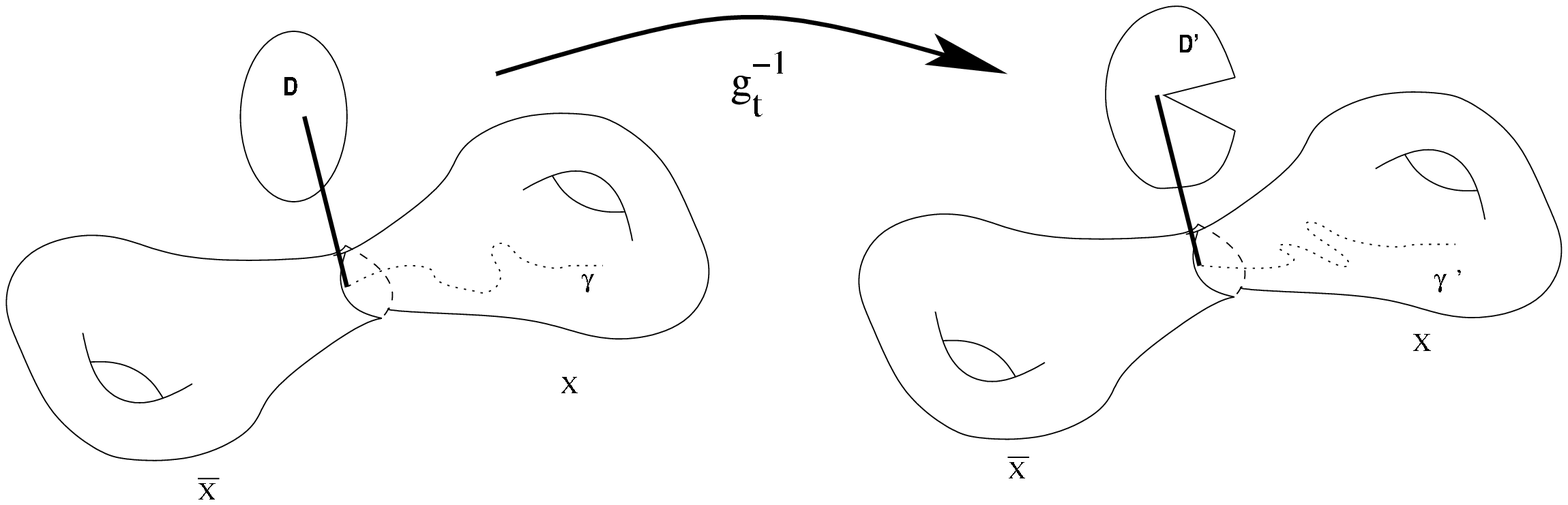}
\mycaption{The Loewner process changes the conformal structure
near the marked point $p$ which is here represented by a deficit
angle in the formal disc.} \label{formaldisc}
\end{center}
\end{figure}

In the special case of the upper half-plane we can simply identify 
the formal (half) disc with the Riemann surface $X=\Hset$ itself. 
In  Sec.~\ref{vira} where we encountered the action of $\Der ~{\cal K}$
through operators ${\cal L}_n$, in particular the Loewner 
process was generated by the second-order hypo-elliptic 
differential operator 
\be
\hat{{\cal H}} \equiv \frac{\kappa}{2} {\cal L}_{-1}^2 -2 {\cal
L}_{-2} ~.
\ee 
In the case of general Riemann surfaces 
this $\Der ~{\cal K}$ action then extends 
to a transitive action on the bundle  
$\widehat{{\cal M}}_{g,1} \longrightarrow {\cal M}_{g,1}$
compatible with the structure group, as was recalled in the previous section.

The CFT analysis leads us to consider sections 
of the line bundle $\Obdle_h$, which is a twisted 
version of the standard determinant bundle defined 
on the moduli space ${\cal M}_{g,1}$. 
By using the above defined projection $\pi$, we can 
construct the pull-back bundle  $\pi^*\Obdle_h$ on 
$\widehat{{\cal M}}_{g,1}$. This bundle carries now a transitive Virasoro 
action, and can be equipped with a flat connection $\nabla=\d + L_{-1}$. 
In this way the $\Der ~{\cal K}$ action is lifted  
to a Virasoro action in the quantum theory. 
The generator of the Loewner process 
\be
\hat{H} \equiv \frac{\kappa}{2} L_{-1}^2 -2 L_{-2}
\ee
should therefore be seen naturally as a map 
\be
\hat{H}:
\Gamma(\pi^* {\Obdle}_h) \longrightarrow \Gamma(\pi^* {\Obdle}_{h+2}) ~.
\ee  
The sections of these pull-back bundles differ from the correlators 
suggested by the CFT analysis only in that they also depend on the 
formal coordinate. This extra structure is just enough to enable us
to equip them with the appropriate Virasoro action and a flat connection. 
  
More precisely, recall that we could consider pull-backs 
by the Loewner mapping $f_t^*$ of conformal forms of 
weight $h$  as stochastic processes $\omega_t \equiv f^*_t \omega$. 
The correlator $\omega_t = \langle \Xope \rangle_{\gamma[0,t]}$
furnishes now an example of such a stochastic process.
These objects are defined as  sections of $\Obdle_h$; When we pull 
these sections back into the bundle  $\pi^*\Obdle_h$ we need to specify
their dependence on the formal coordinate $z \in \Dset$. This can be 
done by requiring that the resulting process  $\pi^* \omega_t$ 
is still a martingale, in the sense that it is annihilated by the action of 
$\hat{H}$. As $\hat{H}$ is of second order, this determines a rank-two 
subbundle $\ker \hat{H} \subset \pi^* {\Obdle}_h$ over ${\cal
M}_{g,1}$. In this way we have been able to 
eliminate  the apriorous dependence on the formal coordinate $z$ 
from the stochastic process
$\pi^* \omega_t$, so that it can really be thought of as an element of 
$\Gamma ({\cal M}_{g,1},\ker \hat{H})$, but have nevertheless been able to 
retain the Virasoro action on it. 

The condition that the pertinent action of 
$\hat{\cal H}$ annihilates the correlator arose also when we 
wanted to  characterise the behaviour of the correlators 
under conditioning; even though we could not present 
an explicit formula, the correlator $\langle \Xope \rangle_{\gamma[0,t]}$
and $\rho^*\langle \Xope \rangle_{\gamma[0,t]}$ both 
were argued to be sections of $\ker \hat{\cal H}$. We see now concretely, 
that the correlator may change under conditioning, but these changes are 
restricted to remain in the properly defined 
rank-two bundle $\ker \hat{H}$. The existence of a flat 
connection on this bundle means that under deformations 
of the complex structure such as in  
Fig.~\ref{cd} the diagram does indeed commute, even though 
the precise holonomies may be more complicated than $Y_t$. The only 
requirement is now that the deformation path in the moduli space is 
contractible. 

\subsection{Probability distributions}

Apart from acting on the stochastic processes $\omega_t$, the
operator $\hat{H}$ also acts on the probability distributions
$\psi(x,t)$ with $z=x+iy$. Where above $\hat{H}\omega_t =0$
implied that $\omega_t$ was a martingale, $\hat{{\cal H}}\psi(x,t)
=0$ implies that the probability distribution is constant in time
$\partial_t \psi =0$. We can therefore also identify the 
suitably normalised positive sections
$\psi\equiv\pi^*{\cal P}(X,x,p)$  of the real line bundle
$\Prob \subset \ker \hat{H}$ as constant in time probability
distributions that solve the diffusion equation in (\ref{genfini})
or, in other words, as zero-energy wave functions of the real
Schr\"odinger equation. 

Let us finally find the explicit form of a generic section 
$\psi \in \Gamma({\cal M}_{g,1}, ~ \Prob)$ of the real subbundle
$\Prob \subset \ker \hat{H}$ of the two-dimensional bundle 
in the case of only one marked point and the formal 
coordinate $z=x+iy$ defined in its local neighbourhood. This amounts to 
finding the real solutions of the
differential equation $\hat{{\cal H}}\psi(z,t) =0$: In polar
coordinates $z=\rho~ \e^{i\varphi}$ they are
\be
\psi(\rho,\varphi) &=& A~ \rho^a \e^{-b \varphi} ~ \cos(a \varphi
+ b \ln \rho + \delta) \label{kerdistr}
\ee
where $A$ and $\delta$ are integration constants and
\be
a &=& \frac{\kappa-4}{2\kappa}  \\
b &=& \frac{\sqrt{32-\kappa^2}}{2\kappa} ~.
\ee
This means that in the range $\kappa \in (0,4)$ the solutions to
this differential equation have a singularity at $z=0$ and are
bounded by $\rho^a$ where $a$ is negative and vanishes precisely
for $\kappa=4$. In particular, when the Hausdorff dimension of the
stochastic process on the Riemann surface reaches  $\dim_H=3/2$ at
$\kappa=4$ and the curve becomes self-touching, the function
$\psi$ reflects this accordingly by ceasing to be suppressed when
$\rho$ tends to infinity.

\section{Discussion}
\label{disc}

We have found that much of the analytical structure of
{SLE}${_{\mathbf\kappa}}$ can be represented in terms of a CFT
with a central charge dictated by $\kappa$. The use of this is the
fact that the CFT treatment extends SLE processes to arbitrary
Riemann surfaces and $\kappa \leq 4$; the drawback is perhaps the
fact that the involved CFT apparatus is still mostly heuristically
defined. Nevertheless, there is a correspondence
between for instance the following objects:

\vspace{3mm}
\begin{center}
\setlength{\extrarowheight}{4pt}
\begin{tabular}{|clccl|}
\hline {\bf SLE}${_{\mathbf\kappa}}$ & {\em Eq.}& & {\bf CFT} &  {\em Eq.} \\
\hline
$\alpha,\lambda$ & (\ref{ak}),(\ref{lk}) & $\longleftrightarrow$ 
& $h,c$ & (\ref{hk}),(\ref{ck}) \\
$P$ & (\ref{girsanov}) & $\longleftrightarrow$ & $ {\cal P}_X(\gamma)$ 
& (\ref{main})  \\
$\psi(x,t)$ &  (\ref{diffdistr})& $\longleftrightarrow$& 
$\pi^*{\cal P}(X,p,z)$ &  (\ref{kerdistr}) \\
$Y_t$ &  (\ref{Y})& $\longleftrightarrow$& $
{ {\rho^*\langle\Xope\rangle_{\gamma[0,t]}}/{\langle\Xope\rangle_{
\gamma[0,t]}}} $ &  (\ref{transfR})\\
$Y_t$ is a martingale & (\ref{Y})  & $\longleftrightarrow$
& $\hat{H} {\langle\Xope\rangle_{\gamma[0,t]}}=0$ & (\ref{singular})  \\
\hline
\end{tabular}
\end{center}
\vspace{4mm}

Most importantly, the measure of
Sec.~\ref{restriction} for clouded paths $\gamma \subset\Xi
\subset X$ enjoys precisely the same covariance properties as the
correlation function ${\cal P}_X(\gamma)$ defined in (\ref{main}) 
when restricted to the upper half-plane. 
Furthermore, the commutativity of the diagram in Figure \ref{cd}
in probability theory is equivalent in CFT to the fact that there
exists a flat connection in $\pi^*{\Obdle}_h$.

As the SLE and the CFT structures appear to be isomorphic it seems
reasonable to conjecture that the probability density
does indeed coincide with the weight constructed in CFT $P(\Xi) =
{\cal P}_X(\gamma)$. The latter construction extends
SLE${_{\mathbf\kappa}}$ to arbitrary Riemann surfaces and to
$\kappa > 8/3$.

\subsubsection*{Acknowledgements}

\begin{sloppypar}
We acknowledge many stimulating discussions with 
M.~Kontsevich, as well as the kind hospitality and support of the 
Institut des Hautes \'Etudes Scientifiques. R.F.~acknowledges 
further the discussions with K.~Linde, V.~Beffara, Y.~Le Jan.  
W.~Werner, he would like to thank having drawn his attention to 
the restriction property very early, 
the discussions on SLE and his help.
\end{sloppypar}

\subsubsection*{Note added}

After the completion of this work the related preprints 
\cite{Bauer:2003kd, maxim} were brought to our attention. 
We would like to thank the Referee for pointing out  
that also Makarov (unpublished) has proposed an extension 
of SLE to non-simply connected geometries.

\end{document}